\documentclass[12pt]{iopart}
\usepackage{graphicx}
\usepackage{dcolumn}
\usepackage{bm}

\def\la{\mathrel{\mathpalette\fun <}}
\def\ga{\mathrel{\mathpalette\fun >}}
\def\fun#1#2{\lower3.6pt\vbox{\baselineskip0pt\lineskip.9pt
  \ialign{$\mathsurround=0pt#1\hfil##\hfil$\crcr#2\crcr\sim\crcr}}}

\begin{document}

\title{Ultra-High Energy Cosmic Ray Nuclei from Individual Magnetized
Sources}
\author{G\"unter Sigl\dag}
\address{\dag GReCO, Institut d'Astrophysique de Paris, C.N.R.S.,
98 bis boulevard Arago, F-75014 Paris, France\\
F\'{e}d\'{e}ration de Recherche Astroparticule et Cosmologie, 
Universit\'{e} Paris 7, 2 place Jussieu, 75251 Paris Cedex 05, France}

\begin{abstract}
We investigate the dependence of composition, spectrum and angular
distributions of ultra-high energy cosmic rays above $10^{19}\,$eV from
individual sources on their magnetization. We find that, especially
for sources within a few megaparsecs from the observer, observable
spectra and composition are severely modified if the source is
surrounded by fields of $\sim10^{-7}\,$G on scales of a few megaparsecs.
Low energy particles diffuse over larger distances during their
energy loss time. This leads to considerable hardening of the spectrum
up to the energy where the loss distance becomes comparable to
the source distance.
Magnetized sources thus have very important consequences
for observations, even if cosmic rays arrive within a few degrees
from the source direction.
At the same time, details in spectra and chemical composition may be
intrinsically unpredictable because they depend on the unknown
magnetic field structure.
If primaries are predominantly nuclei of atomic mass
$A$ accelerated up to a maximum energy $E_{\rm max}$ with spectra
not much softer than $E^{-2}$, secondary protons from
photo-disintegration can produce a conspicuous peak in the spectrum
at energy $\simeq E_{\rm max}/A$. A related feature appears in the
average mass dependence on energy. 
\end{abstract}



\vskip1cm

\noindent version published as JCAP08(2004)012,
Copyright Institute of Physics and IOP Publishing Limited 2004.

\maketitle

\section{Introduction}
The origin of ultra-high energy cosmic rays (UHECR) above $10^{19}\,$eV
($=10\,$EeV) is a mystery since many years~\cite{reviews,school}. Several
next-generation experiments, most notably the Pierre Auger
experiment now under construction~\cite{auger} and the EUSO
project~\cite{euso} are now trying to solve this mystery.

Although statistically meaningful information about the UHECR energy
spectrum and arrival direction distribution has been accumulated, no
conclusive picture for the nature and distribution of the sources
emerges naturally from the data. There is on the one hand the approximate
isotropic arrival direction distribution~\cite{bm} which indicates that we are
observing a large number of weak or distant sources. On the other hand,
there are also indications which point more towards a small number of
local and therefore bright sources, especially at the highest energies:
First, the AGASA ground array claims statistically significant multi-plets of
events from the same directions within a few degrees~\cite{teshima1,bm},
although this is controversial~\cite{fw} and has not been seen so far
by the fluorescence experiment HiRes~\cite{finley}.
The spectrum of this clustered component is $\propto E^{-1.8}$ and thus
much harder than the total spectrum~\cite{teshima1}.
Second, nucleons above $\simeq70\,$EeV suffer heavy energy losses due to
photo-pion production on the cosmic microwave background
--- the Greisen-Zatsepin-Kuzmin (GZK) effect~\cite{gzk} ---
which limits the distance to possible sources to less than
$\simeq100\,$Mpc~\cite{stecker}. Heavy nuclei at these energies
are photo-disintegrated in the cosmic microwave background within a
few Mpc~\cite{heavy}. For a uniform source distribution
this would predict a ``GZK cutoff'', a drop in the spectrum.
However, the existence of this ``cutoff'' is not established yet
from the observations~\cite{bergman}.

The picture is further complicated by the likely presence of large
scale extra-galactic magnetic fields (EGMF) that will lead to deflection
of any charged UHECR component.
Magnetic fields are omnipresent in the Universe, but their
true origin is still unclear~\cite{bt_review}. Magnetic fields
in galaxies are observed with typical strengths of a few
micro Gauss, but there are also some indications for fields correlated
with larger structures such as galaxy clusters~\cite{bo_review}.
Magnetic fields as strong as
$\simeq 1 \mu G$ in sheets and filaments of the large scale galaxy
distribution, such as in our Local Supercluster, are compatible with
existing upper limits on Faraday rotation~\cite{bo_review,ryu}.
It is also possible that fossil cocoons of former radio galaxies,
so called radio ghosts, contribute significantly to the isotropization
of UHECR arrival directions~\cite{mte}.

Only recently attempts have been made
to simulate UHECR propagation in a realistically structured and
magnetized universe based on large scale structure
simulations~\cite{sme,dolag}. These two simulations used
different models for the EGMF: Whereas seed fields were continuously
injected at shocks in Ref.~\cite{sme}, Ref.~\cite{dolag} started
from uniform seed fields. In both cases, the seed field strength
had to be normalized to reproduce observed rotation measures.
Interestingly, the EGMF in Ref.~\cite{sme} is considerably more
extended and leads to much larger average deflections than in
Ref.~\cite{dolag}. The main reason is probably due to the different
EGMF models. This suggests that the influence of EGMF on UHECR
propagation is currently hard to quantify. The two works agree,
however, at least in the following: First, the observer is most likely
situated in a region with relatively weak EGMF, $\la10^{-9}\,$G.
In Ref.~\cite{sme} which used an unconstrained large scale structure
simulation, this was deduced from comparison of predicted
and observed UHECR anisotropies, whereas in Ref.~\cite{dolag}
this followed from their constrained simulation which is supposed
to lead to an approximate model of our local extra-galactic environment.
Second, the fields are not strong and extended enough to allow an
interpretation of the large scale isotropy of UHECR arrival directions
by only one nearby source, as suggested in Ref.~\cite{farrar}.
Finally, both these works were restricted to nucleons. Heavy
nuclei may further complicate the issue and lead to large deflection
even in the EGMF scenario of Ref.~\cite{dolag}.

In the present paper we study UHECR spectra, composition and angular
distribution from individual discrete sources injecting protons
or nuclei into a highly structured EGMF which is significant
only in the neighborhood of the source. This is different from the
more idealized Monte Carlo simulations for individual sources
in Refs.~\cite{slo,tanco,slb,bils,sse} in the following points: The
EGMF in these earlier simulations was idealized by a Kolmogorov spectrum with
a possible coherent component, with a field energy density either
following a Gaussian profile~\cite{slb,sse} or being
constant~\cite{slo,bils}, or as organized in spatial cells with a
given coherence length and a strength depending as a power law on the
local density~\cite{tanco}. Furthermore, the
observer was immersed in fields of similar strength as the source,
and these simulations were restricted to nucleons.

In the context of a highly structured EGMF it was already found that
the auto-correlation
function at degree scales decreases with source magnetization
and can thus be used as a signature for magnetic fields~\cite{letter,sme}.
In the present paper we will find that even if off-sets of arrival directions
from the source direction are moderate, spectra and composition
are considerably modified if the source is surrounded by EGMF
of the order $10^{-7}\,$G on scales of a few Mpc.

In Sect.~2 we describe our simulations, in Sect.~3 the results are
presented, and we conclude in Sect.~4.

\section{Simulations}

\begin{figure}[ht]
\includegraphics[width=0.5\textwidth,clip=true]{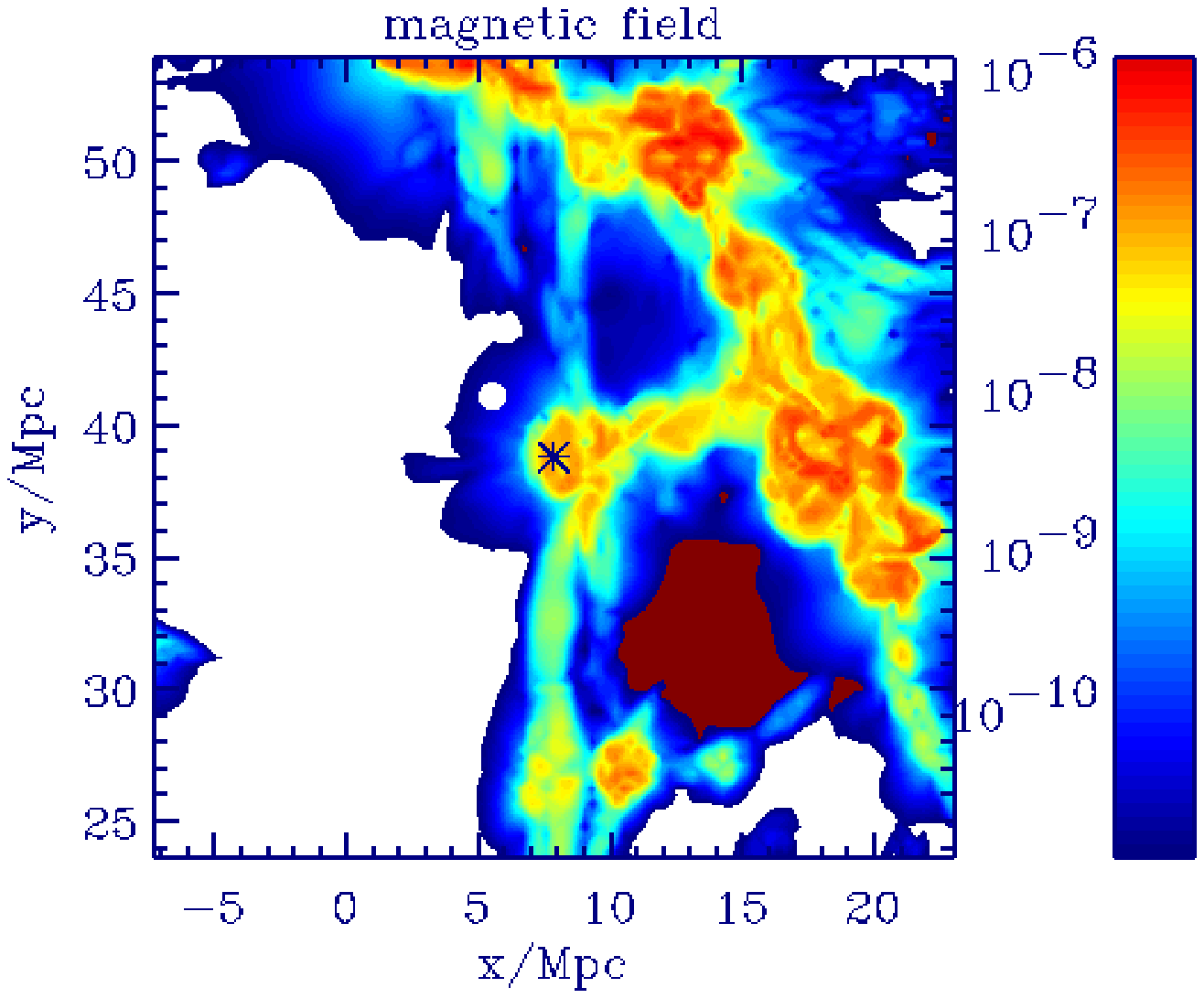}
\includegraphics[width=0.5\textwidth,clip=true]{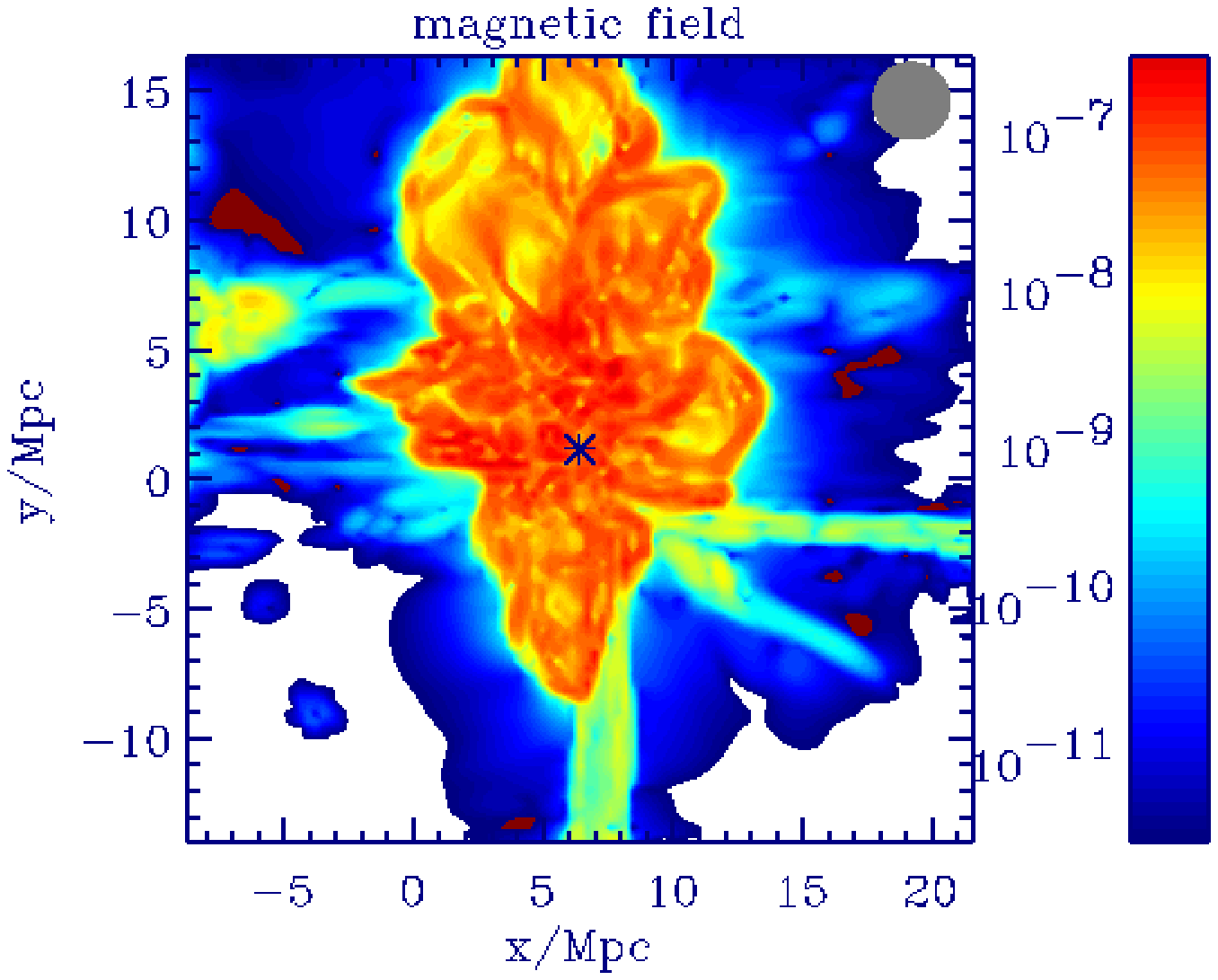}
\caption[...]{Log-scale two-dimensional cuts through magnetic field
total strength (color scale, in Gauss) with source position indicated
by the asterisk in the center. Left panel: Scenario ``M82''. The observer
is represented by the white disk to the left and above the source.
Right panel: Scenario ``Virgo''. The observer is represented
by the grey disk in the upper right corner.}
\label{fig1}
\end{figure}

We will specifically study two cases where the location of the
observer and source are chosen from the large scale structure simulation
used in Ref.~\cite{sme}, see Ref.~\cite{miniati}. In the first
case the source is at a distance of 3.3 Mpc, similar to the
starburst galaxy M82 or the radio galaxy Centaurus A. The field
strength at the observer is
$\simeq3\times10^{-11}\,$G, whereas the source is immersed in
fields of $\sim10^{-7}\,$G. We will denote this scenario by ``M82''.
In the second case, the source is immersed
in a galaxy cluster at 18.6 Mpc distance, reminiscent of the Virgo
cluster. The field strength at the observer is $\simeq10^{-11}\,$G,
whereas the source is immersed in fields of a few $10^{-7}\,$G on
scales of several Mpc. We will denote this scenario by ``Virgo''.
These two cases are visualized in Fig.~\ref{fig1}.
We consider the steady state situation which amounts to assuming that
these sources are active on time scales at least as long as
the propagation time. In this case observables are time independent.

The trajectory simulations are performed as described in Ref.~\cite{sme}.
These simulations were generalized to include heavy nuclei and follow
all secondary trajectories produced by photo-disintegration reactions, in the
way described in Ref.~\cite{bils}. Interactions taken into account
thus include pair production by protons, pion production, and
photo-disintegration on the combined cosmic microwave,
infrared/optical and radio photon backgrounds. For the latter,
standard estimates were used, see Ref.~\cite{bils} for details.

For simplicity, we will assume that nuclei are accelerated with a
spectrum $\propto E^{-2}$ up to a fixed maximal energy
$E_{\rm max}=4\times10^{21}\,$eV, and we will restrict ourselves to
either proton or iron primaries at injection. These are reasonable
assumptions for radio galaxies~\cite{rachen}. We note that active
galaxies are likely to accelerate heavy nuclei to ultra-high
energies, see Refs~\cite{sme,agrs} for discussions.

The observer is modeled as a sphere and a trajectory is accepted and
observationally relevant
quantities are saved every time it crosses this sphere. For each
scenario $10^6$ arriving trajectories were simulated.
Galactic magnetic fields lead to deflections of maximally a
few tens of degrees above $10^{19}\,$eV and can therefore be neglected
as long as one is mainly interested in spectra and chemical composition.

\section{Results}

\begin{figure}[ht]
\includegraphics[width=0.5\textwidth,clip=true]{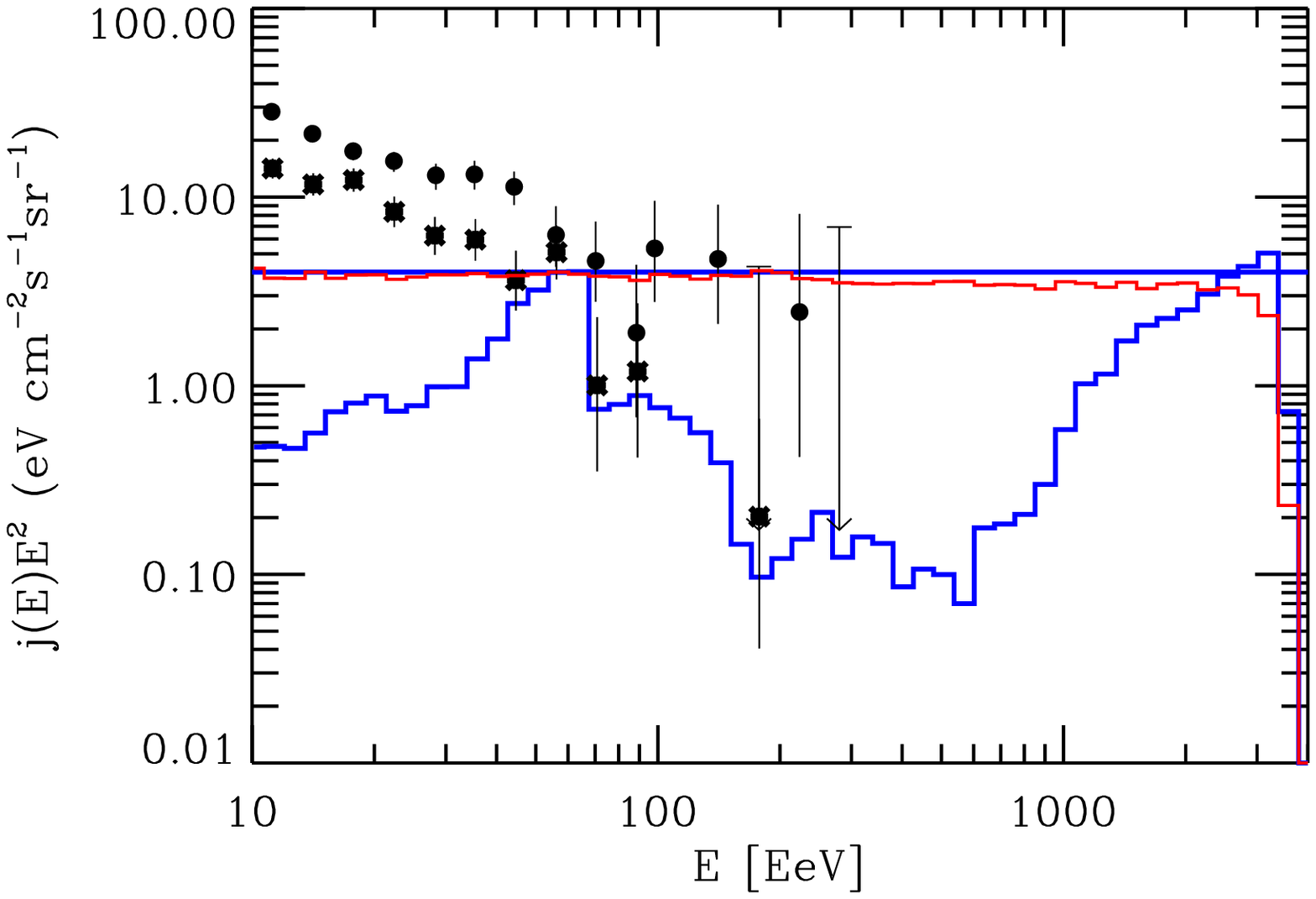}
\includegraphics[width=0.5\textwidth,clip=true]{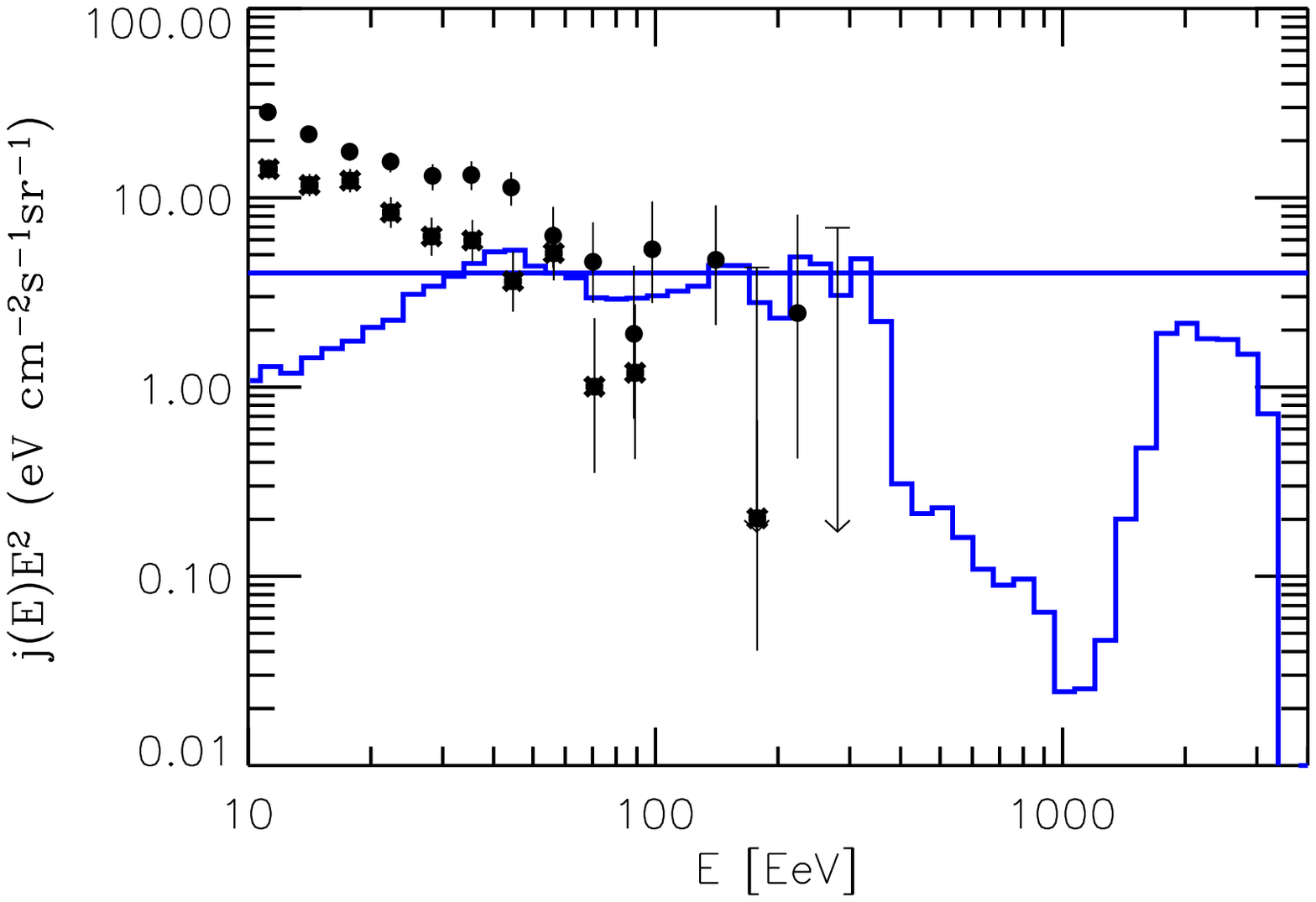}
\caption[...]{Steady state all-particle spectra predicted by scenario
``M82'' for iron primaries with injection spectrum $\propto E^{-2.0}$ up to
$4\times10^{21}\,$eV. The blue curves are for the
EGMF surrounding the source shown in Fig.~\ref{fig1}, left panel,
and the red curve is without EGMF.
Shown for comparison are the solid angle integrated
AGASA~\cite{agasa} (dots) and HiRes-I~\cite{hires} (stars) data.
The solid straight line marks the injection spectrum. The right
panel is for a source off-set by about 1 Mpc from the position
shown in Fig.~\ref{fig1}, left panel. All fluxes have been normalized
at 60 EeV.}
\label{fig2}
\end{figure}

\begin{figure}[ht]
\includegraphics[width=0.5\textwidth,clip=true]{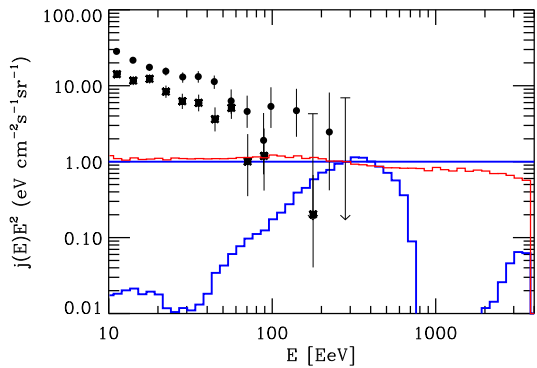}
\caption[...]{Same as Fig.~\ref{fig2}, left panel, but for proton
primaries. Here the fluxes have been normalized at 300 EeV.}
\label{fig3}
\end{figure}

Fig.~\ref{fig2} compares the solid angle integrated spectra predicted by
scenario ``M82'' for iron primaries, with and without EGMF. The left
panel corresponds to the source position shown in Fig.~\ref{fig1},
left panel, whereas the right panel is for a slightly
different source position for comparison. Fig.~\ref{fig3} shows the
same for proton primaries for the source position shown in Fig.~\ref{fig1},
left panel. The normalizations are chosen such that this individual source
can contribute at least partly to the measured UHECR flux at the
highest energies. At lower energies cosmological sources are
likely to dominate the observed flux~\cite{sme}.

\begin{figure}[ht]
\includegraphics[width=0.5\textwidth,clip=true]{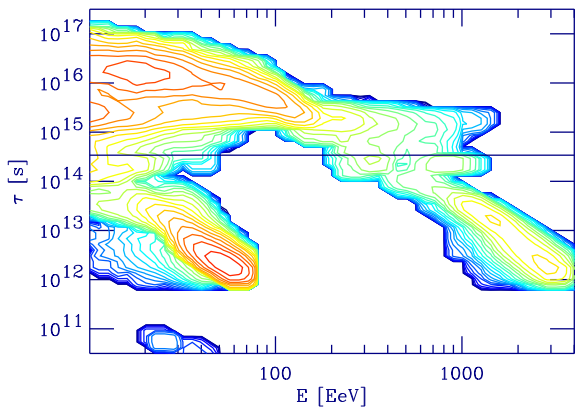}
\caption[...]{The distribution of time delays and arrival energies
for the scenario ``M82'' for iron primaries, corresponding to
Fig.~\ref{fig2}, left panel. The horizontal line
corresponds to the rectilinear propagation time.}
\label{fig4}
\end{figure}

\begin{figure}[ht]
\includegraphics[width=0.5\textwidth,clip=true]{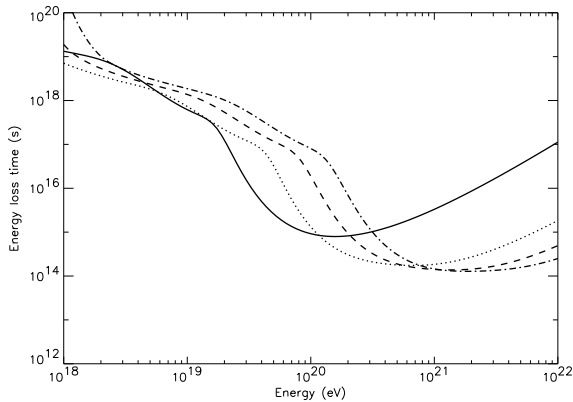}
\caption[...]{The energy loss time as a function of energy for
photo-disintegration on the combined cosmic microwave, infra-red
and radio backgrounds. The solid line is for Helium nuclei,
the dotted line for Carbon, the dashed line for Silicon, and the
dash-dotted line for Iron. From Ref.~\cite{bils}.}
\label{fig5}
\end{figure}

There are several notable features for the
nearby source ``M82'' which we will discuss in turn: It is first
of all apparent that whereas without fields the spectrum is
only slightly modified due to the small source distance, the
magnetized case leads to a strong modification of the spectrum.
This can be understood by comparing the time delay due to
propagation in the EGMF, shown in Fig.~\ref{fig4}, with typical
energy loss times, shown in Fig.~\ref{fig5}. The two time scales
can indeed be comparable.

Furthermore, the spectra shown in Figs.~\ref{fig2} and~\ref{fig3}
refer to the steady state situation relevant as long as
the source emission characteristics do not change considerably
on the time scale of typical time delays. This may cease to
be realistic at low energies where time delays reach several
hundred million years according to Fig.~\ref{fig4}. Sources
which were inactive at such times before the present would
thus not be visible around $10^{19}\,$eV today, because such
UHECR would not have reached us yet. This would produce a
strong cut-off at low energies.

We are here dealing with highly structured EGMF concentrated
on scales of a few Mpc, with a typical r.m.s. strength $B$
and coherence length $L_c$. As long as the latter is not much
smaller than the length scale over which the EGMF extends,
the transition from rectilinear propagation
to diffusion roughly occurs when the Larmor radius
$r_L(E)\simeq E/(ZeB)$ becomes comparable to $L_c$, or at
\begin{equation}
  E_t\simeq 10^{20}\,Z\left(\frac{B}{10^{-7}\,{\rm G}}\right)
  \left(\frac{L_c}{1\,{\rm Mpc}}\right)\,{\rm eV}\,.\label{Et}
\end{equation}
For $E\la E_t$ the time delay
scales asymptotically as $\tau(E)\propto E^{-1/3}$, whereas in
the rectilinear regime $\tau(E)\propto E^{-2}$~\cite{slb}.
According to Eq.~(\ref{Et}), for light nuclei and $L_c\sim100\,$kpc
this transition occurs indeed between $10^{19}\,$eV and $10^{20}\,$eV,
consistent with what is seen in Fig.~\ref{fig4} for the branch
with large delay times. This is consistent with the simulations
of Ref.~\cite{dolag} which has weaker and less extended fields
than used in our simulations, but where in typical galaxy clusters
protons diffuse nevertheless up to several $10^{19}\,$eV, as discussed
in Ref.~\cite{rgd}.

Another conspicuous feature in case of iron primaries is a rise
of the spectrum towards lower energies at $E\simeq4-6\times10^{19}\,$eV,
see Fig.~\ref{fig2}, whose strength depends, however, considerably
on the detailed EGMF realization. This energy roughly
coincides with $E_{\rm max}/A$ where $A=56$ for iron. It is
thus given by the maximal energies of nucleons produced as
secondaries by iron photo-disintegration. This is confirmed by the
average atomic mass as a function of energy shown in Fig.~\ref{fig6}
which shows a sharp drop at the same energy $E\simeq6\times10^{19}\,$eV
due to these nucleons. Since such nucleons at energy $E$ are produced
by the flux of nuclei at energy $A E$, their effect is visible
only for injection spectra not much softer than $E^{-2}$.
The nucleon threshold is visible also in a branch
of events with delay times $\tau\sim10^{12}\,$s around $E\sim60\,$EeV
in Fig.~\ref{fig4}. This branch is an almost identical copy of
the analogous heavy nuclei branch with the same delay times at
energies $\simeq 56$ times higher. It is thus due to photo-disintegration
of these nuclei into nucleons away from the source where the weak
magnetic field hardly adds any time delay.

Finally, again for iron primaries, the spectrum is considerably
hardened in two energy intervals stretching over about half
an order of magnitude, at $E\la6\times10^{19}\,$eV
and at $E\la E_{\rm max}$, see Fig.~\ref{fig2}, left panel. These
two intervals are off-set from each other by roughly the charge of
the iron primaries, $Z=26$. A similar hardening is
seen for proton primaries in the observed spectrum below
$\simeq3\times10^{20}\,$eV, see Fig.~\ref{fig3}.
The hardening in these energy intervals can be interpreted by
diffusive effects in terms of approximate analytical terms
as follows:

\begin{figure}[ht]
\includegraphics[width=0.5\textwidth,clip=true]{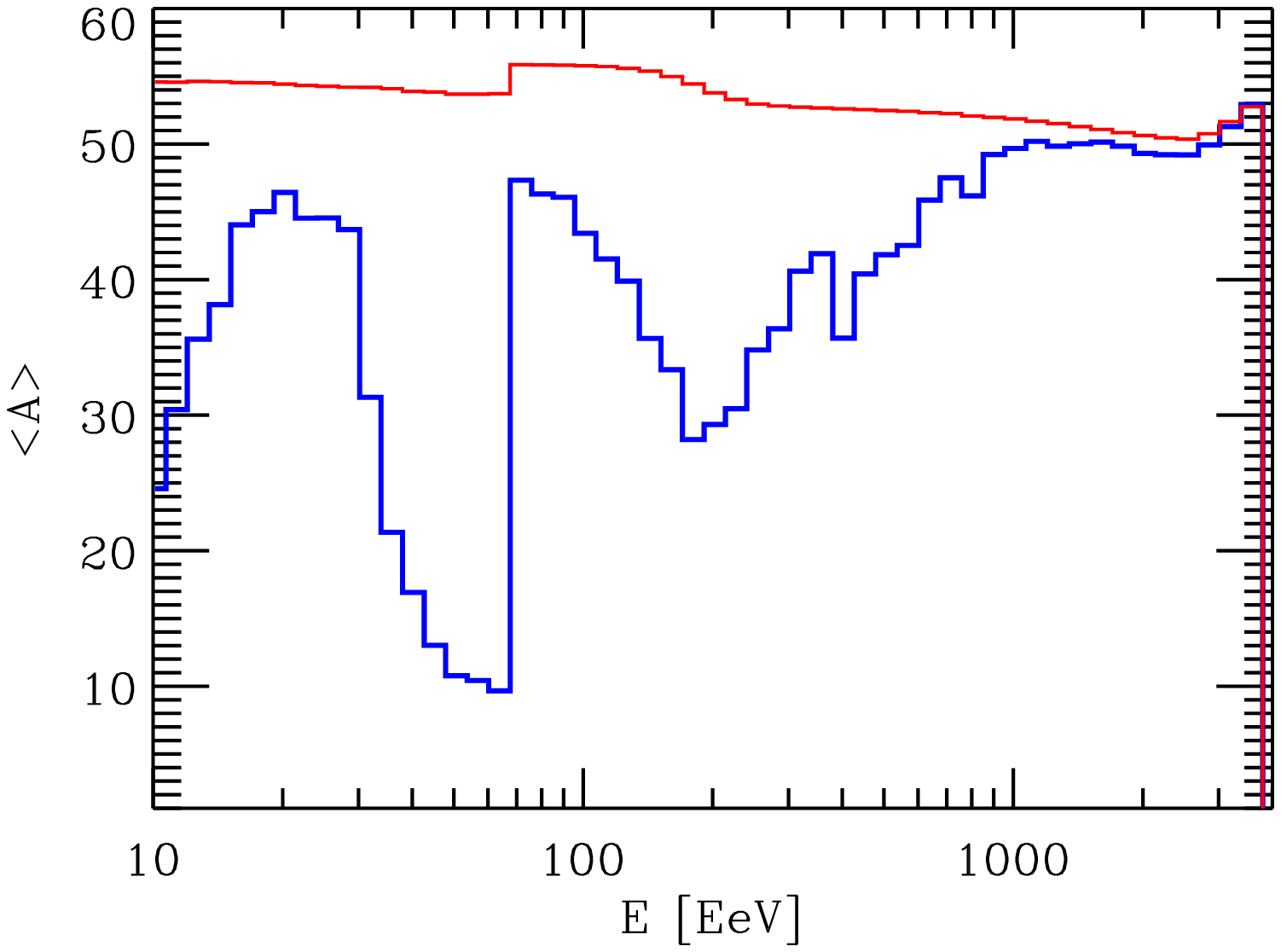}
\includegraphics[width=0.5\textwidth,clip=true]{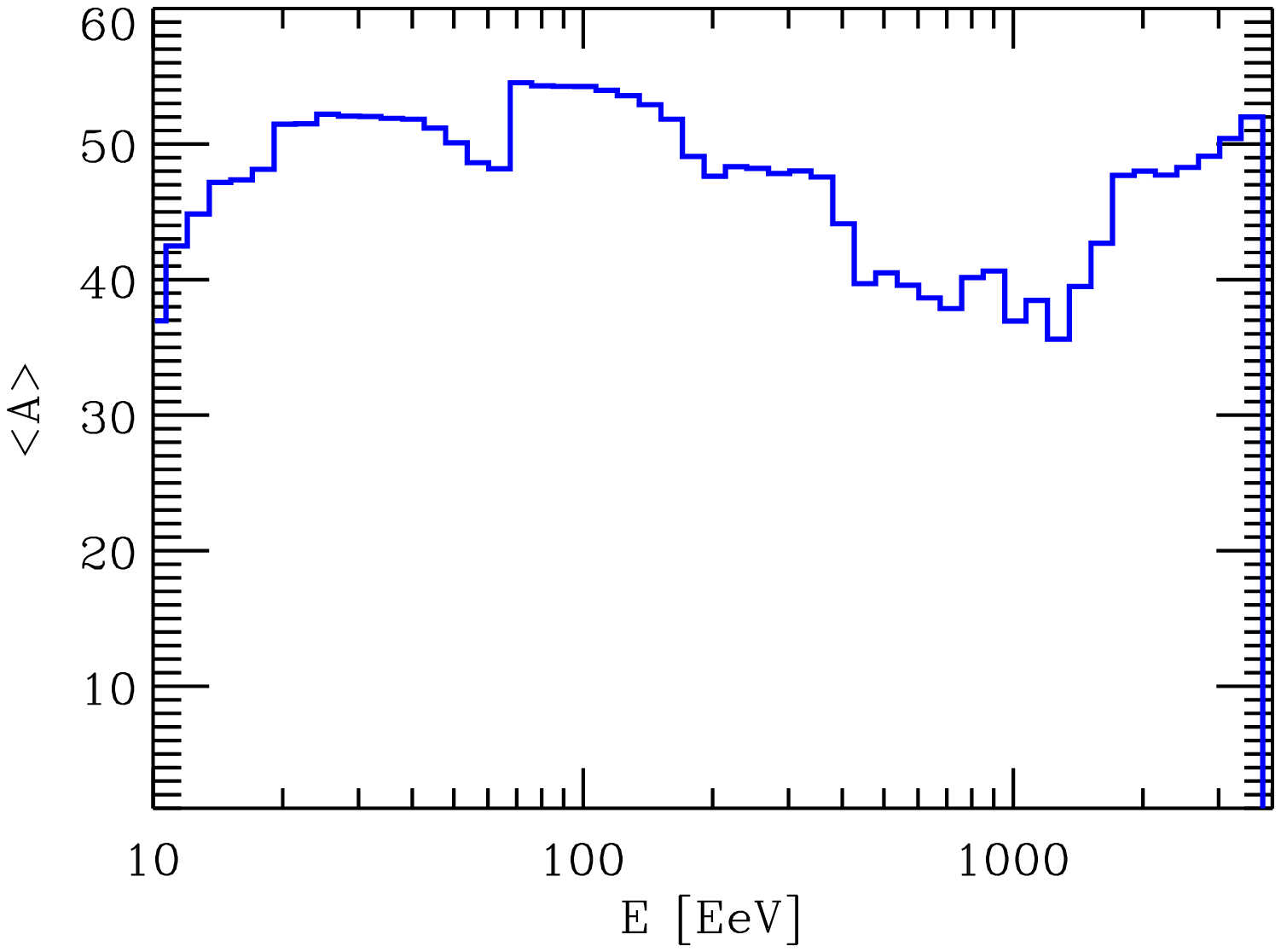}
\caption[...]{Average atomic mass of observed UHECR as a function
of energy predicted by scenario ``M82'' for iron primaries. The blue
curves are for the
EGMF surrounding the source as in Fig.~\ref{fig1}, left panel,
and the red curve is without EGMF. The two panels correspond
exactly to the two panels of Fig.~\ref{fig2} and demonstrate the
dependence on source position.}
\label{fig6}
\end{figure}

In the homogeneous case the diffusion - energy loss
equation characterized by an energy dependent diffusion coefficient
$D(E)$ and (continuous) energy losses, $dE/dt=-E/\tau_{\rm loss}(E)$,
has the solution~\cite{B91,aloisio,slb}
\begin{equation}
j(E)=\frac{1}{4\pi}
\int_{E}^{+\infty}\frac{dE^\prime}{E}\,\frac{Q(E^\prime)\tau_{\rm loss}(E)}
{\left[4\pi\lambda(E,E^\prime)^{2}\right]^{3/2}}\,
\exp\left[-\frac{d^2}{4\lambda(E,E^\prime)^2}\right]\,,\label{Eq_diff_inf}
\end{equation}
for the flux $j(E)$. Here, $Q(E)$ is the injection spectrum of the
discrete source at distance $d$ and
\begin{equation}
\lambda(E_1,E_2)\equiv \left[\int_{E_1}^{E_2}d\ln E^\prime\,
\tau_{\rm loss}(E^\prime)D(E^\prime)\right]^{1/2}\,,\label{dEloss}
\end{equation}
denotes an effective path length against energy losses.
The diffusion coefficient scales as $D(E)\propto E^\alpha$
with $\alpha\la1$~\cite{slb}, and $\tau_{\rm loss}(E)$ grows quickly
with decreasing energy, see Fig.~\ref{fig5}. As a result, $\lambda(E,E^\prime)$
increases with decreasing $E$. This gives rise to a change of
slope in the flux Eq.~(\ref{Eq_diff_inf}), at an energy $E_d$ where
$\lambda(E,E^\prime)\simeq d$: At energies $E\ga E_d$ where
$\lambda(E,E^\prime)\la d$, the flux starts to become
exponentially suppressed according to Eq.~(\ref{Eq_diff_inf})
which corresponds to the usual GZK-type cutoff. At energies $E\la E_d$
the spectrum is modified by $\tau_{\rm loss}(E)/\lambda(E,E^\prime)^3\propto
\tau_{\rm loss}(E)^{-1/2}D(E)^{-3/2}$ relative to the injection spectrum.
This factor is the ratio of the accumulation time $\tau_{\rm loss}(E)$ and the
volume $\lambda(E,E^\prime)^3$ over which UHECR of energy $E$ are
distributed. When multiplied with the injection spectrum $Q(E^\prime)$
in Eq.~(\ref{Eq_diff_inf}), it thus gives the UHECR density.
In an EGMF with uniform statistical properties,
$D(E)^{-3/2}\propto E^{-3\alpha/2}$ can increase faster with
decreasing energy $E$ than $\tau_{\rm loss}(E)^{-1/2}$ decreases.
This can lead to a {\it steepening} of the spectrum as observed,
e.g., in the simulations of Refs.~\cite{slb,bils,lsb} and discussed
analytically in Ref.~\cite{aloisio}. In contrast, in our inhomogeneous
scenarios the effective diffusion coefficient $\overline{D(E)}$
depends more weakly on $E$ because lower energy particles diffuse
to larger distances from the source during their larger energy loss
time. The correspondingly smaller EGMF partly compensates the decrease
of $\overline{D(E)}$ with $E$. As a consequence, the factor
$\tau_{\rm loss}(E)^{-1/2}\overline{D(E)}^{-3/2}$ in general leads
to suppression with decreasing energy.

Since according to Fig.~\ref{fig5} the minimal $\tau_{\rm loss}(E)$
is comparable to the source distance $d$ in our scenarios, and
since the transition from diffusion to straight line propagation occurs
when $D(E)\sim r_L(E)\sim L_c$ which, as $d$, is of order Mpc, the
energy $E_d$ is also roughly equal to the energy $E_t$ defined in
Eq.~(\ref{Et}).
This provides a rough explanation of the transition from diffusive
hardening at low energies to softening at higher energies seen in
the spectra of Figs.~\ref{fig2}, Fig.~\ref{fig3}, and Fig.~\ref{fig8}
below. It should be
kept in mind, however, that our scenarios are highly inhomogeneous,
the diffusion approximation strictly speaking does
not apply far from the source, and these arguments can thus be considered
only as approximate at best. Monte Carlo simulations are indispensable
in this situation.

In order to test how much the results depend on the detailed realization
of EGMF and source, we also performed simulations in which the exact
source position is varied within 1-2 Mpc within the general structures
shown in Fig.~\ref{fig1}. We find that the general tendency to spectral
hardening at energies where UHECR diffuse in the vicinity of the source
is always seen. However, the detailed shape of the predicted observable
spectra depends considerably on the concrete structure of the EGMF.
This is due to magnetic lensing~\cite{hmrs} which leads to flux
enhancements along specific propagation paths at specific energies,
especially if there are field lines connecting source and observer.
This is analogous to the direction dependence of the simulated
nucleon spectrum from a source at the center of a magnetic slab
obtained in Ref.~\cite{sse}. Due to the unknown magnetic field structure
these spectral details are in general unpredictable. This is demonstrated
by comparing the two panels in Fig.~\ref{fig2} for iron primaries.
The same is true for the detailed angular distributions.

\begin{figure}[ht]
\includegraphics[width=0.5\textwidth,clip=true]{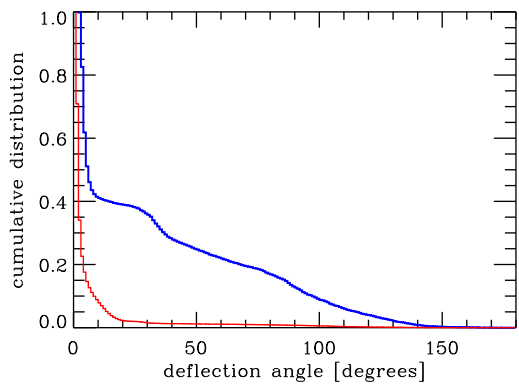}
\caption[...]{The cumulative distribution of arrival direction
off-sets from the source direction for UHECR above $4\times10^{19}\,$eV
in scenario ``M82'' for iron primaries, corresponding to
Fig.~\ref{fig2}, left panel (blue curve) and proton
primaries, corresponding to Fig.~\ref{fig3} (red curve).}
\label{fig7}
\end{figure}

Fig.~\ref{fig6} shows the average atomic mass of observed UHECR
as a function of energy for iron primaries. The nucleon spike
towards low $\left\langle A\right\rangle$ at $E\simeq6\times10^{19}\,$eV
is clearly visible, as discussed above. Below that energy there
appears, however, also a relatively heavy component which survived
excessive photo-disintegration. At all energies, the EGMF reduces the average
atomic number due to increased photo-disintegration compared to undeflected
propagation. Apart from this and the location of the nucleon spike,
the detailed mass distribution is, however, quite sensitive to
the EGMF structure, as seen by comparing the two panels in Fig.~\ref{fig6}.

Fig.~\ref{fig7} shows the cumulative distribution
of arrival direction off-sets from the source direction for the
same ``M82'' scenario. Clearly, deflections are larger for
iron primaries, whereas for proton primaries they are only
a few degrees.

\begin{table}[ht]
\caption[...]{\label{tab1} Approximate source emission power above
$\sim10^{19}\,$eV in erg$/$s corresponding to the normalizations of the
observed spectra in Figs.~\ref{fig2}, left panel, \ref{fig3},
and~\ref{fig8}.}
\begin{indented}
\lineup
\item[]\begin{tabular}{@{}*{3}{l}}
\br
 & with EGMF & no EGMF \\
\mr
M82, iron primaries & $2.8\times10^{40}$ &
$6.6\times10^{40}$ \\
M82, proton primaries & $3.8\times10^{39}$ &
$1.7\times10^{40}$ \\
Virgo, iron primaries & $3.9\times10^{42}$ &
$2.4\times10^{42}$ \\
Virgo, proton primaries & $2.0\times10^{42}$ &
$1.7\times10^{42}$ \\
\br
\end{tabular}
\end{indented}
\end{table}

\begin{figure}[ht]
\includegraphics[width=0.5\textwidth,clip=true]{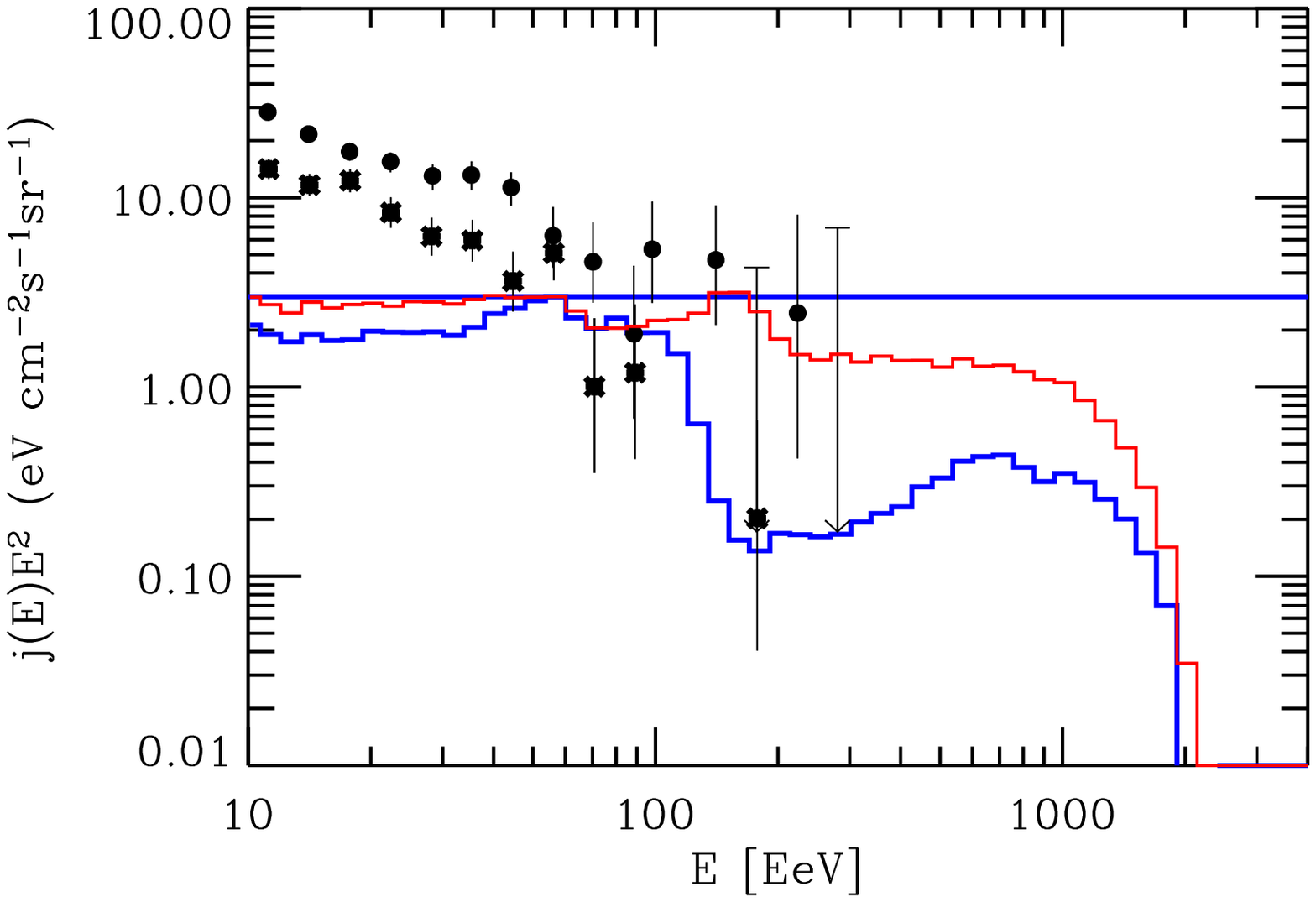}
\includegraphics[width=0.5\textwidth,clip=true]{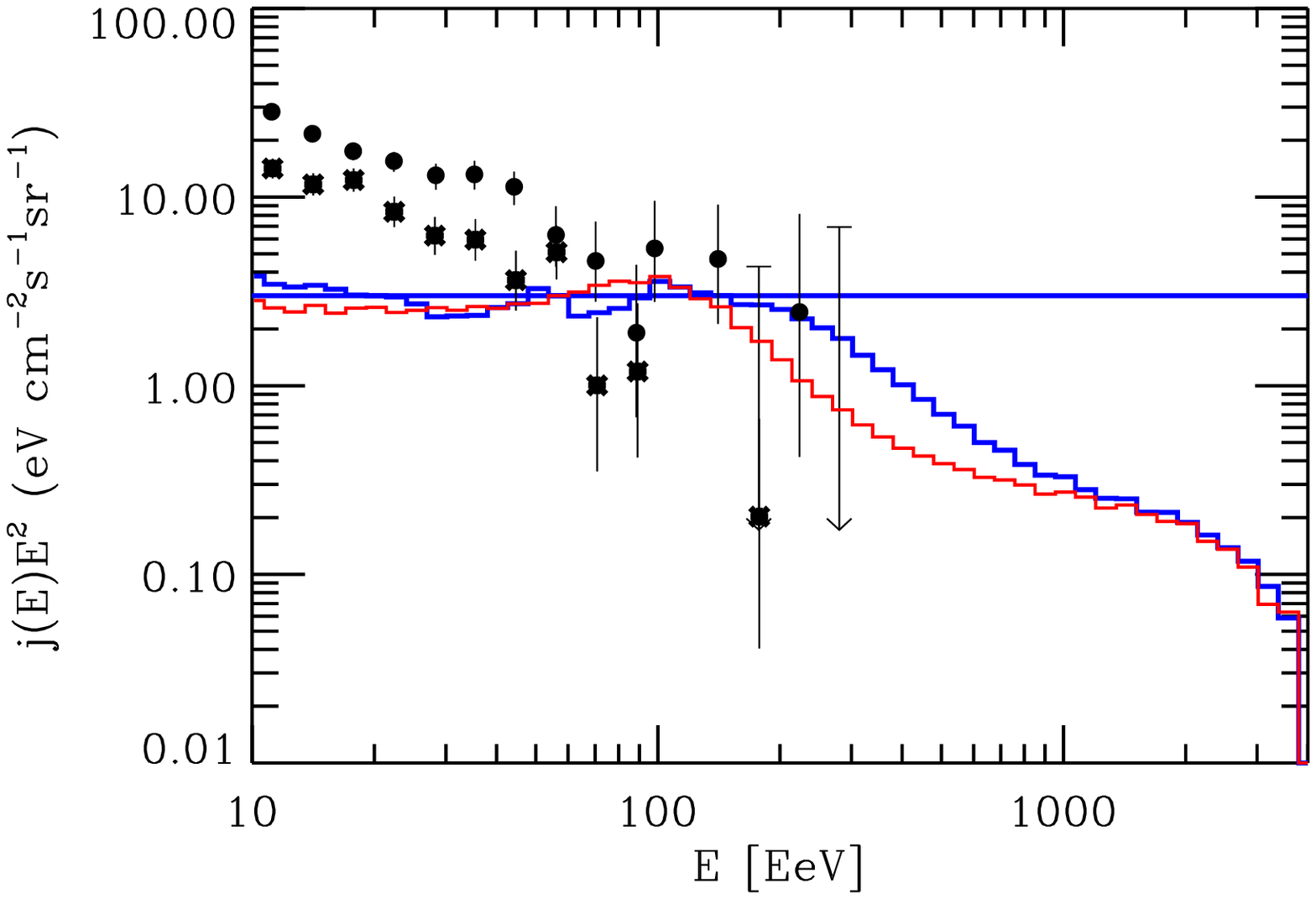}
\caption[...]{Same as Figs.~\ref{fig2} and~\ref{fig3} but for
scenario ``Virgo'', shown in the right panel of Fig.~\ref{fig1}.
The left and right panels are for iron and
proton primaries, respectively. All fluxes have been normalized at
60 EeV.}
\label{fig8}
\end{figure}

\begin{figure}[ht]
\includegraphics[width=0.5\textwidth,clip=true]{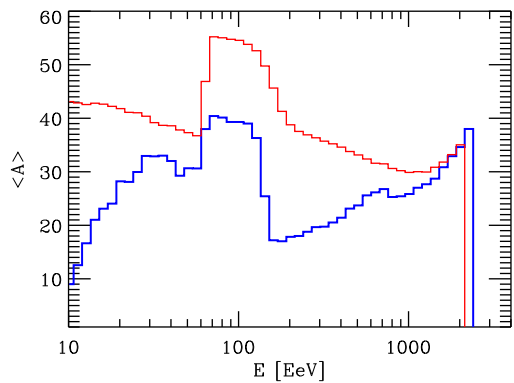}
\caption[...]{Same as Fig.~\ref{fig6}, but for scenario ``Virgo''.}
\label{fig9}
\end{figure}

\begin{figure}[ht]
\includegraphics[width=0.5\textwidth,clip=true]{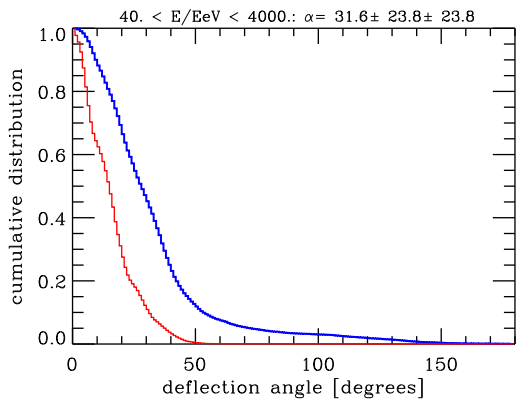}
\caption[...]{Same as Fig.~\ref{fig7}, but for scenario ``Virgo''.
This corresponds to the cases shown in Fig.~\ref{fig8}.}
\label{fig10}
\end{figure}

We can also extract from the simulations the necessary source
power corresponding to a particular normalization of the
observed spectrum. For the normalizations shown in Figs.~\ref{fig2},
\ref{fig3}, and~\ref{fig8} below the source power is given in
Tab.~\ref{tab1}. The required power is not strongly changed
by the presence of the EGMF because we have normalized at
the energy where the energy flux, which is proportional to the y-axis
in Figs.~\ref{fig2}, \ref{fig3} and~\ref{fig8}, is maximal. Note
that for a given normalization of the observed spectrum, iron
primaries require a somewhat higher injection power because a
higher fraction of the injected energy is degraded below
$10^{19}\,$eV during propagation due to spallation. For the case
of ``M82'' most of the difference in required injection power
between proton and iron primaries is due to the difference in
normalization of the observed spectrum in these two cases,
compare Figs.~\ref{fig2} and~\ref{fig3}. Finally, the unknown
EGMF structure translates into uncertainties in the required
power of about an order of magnitude. Given these uncertainties of the
injection spectrum and its continuation below $10^{19}\,$eV,
these values are consistent with the UHECR emission power
of active galaxies discussed in this context~\cite{rachen}, such
as for M87, the main radio galaxy in the Virgo cluster~\cite{protheroe}.

Figs.~\ref{fig8}, \ref{fig9}, and \ref{fig10} show spectra,
atomic mass distribution, and the cumulative deflection angle
distribution, respectively, for the ``Virgo'' scenario. For
direct comparison with the ``M82'' scenario, the same injection
parameters have been assumed.
The effects of the EGMF surrounding the source, although more extended
than in the scenario ``M82'', are here relatively smaller.
This is because the delay time relative to the straight line
propagation time now only reaches a maximum of $\sim10$ at
$\simeq10^{19}\,$eV, compared to $\sim100$ in the scenario ``M82''.
The EGMF effects are, however, still significant and show the
same generic features as discussed for the case ``M82''. In fact,
for the same injection spectrum, at low energies, the suppression
of the average atomic mass due to increased photo-disintegration by propagation
in the EGMF is more severe than for the more nearby source
``M82'', compare Figs.~\ref{fig6} and~\ref{fig9}. This is because
due to the larger propagation distance, fewer nuclei survive
spallation.

Note that the conventional pile-up in the spectrum around 100 EeV is
clearly visible in the case of proton primaries without EGMF,
see Figs.~\ref{fig3} and~\ref{fig8}, right panels.

\begin{figure}[ht]
\includegraphics[width=0.5\textwidth,clip=true]{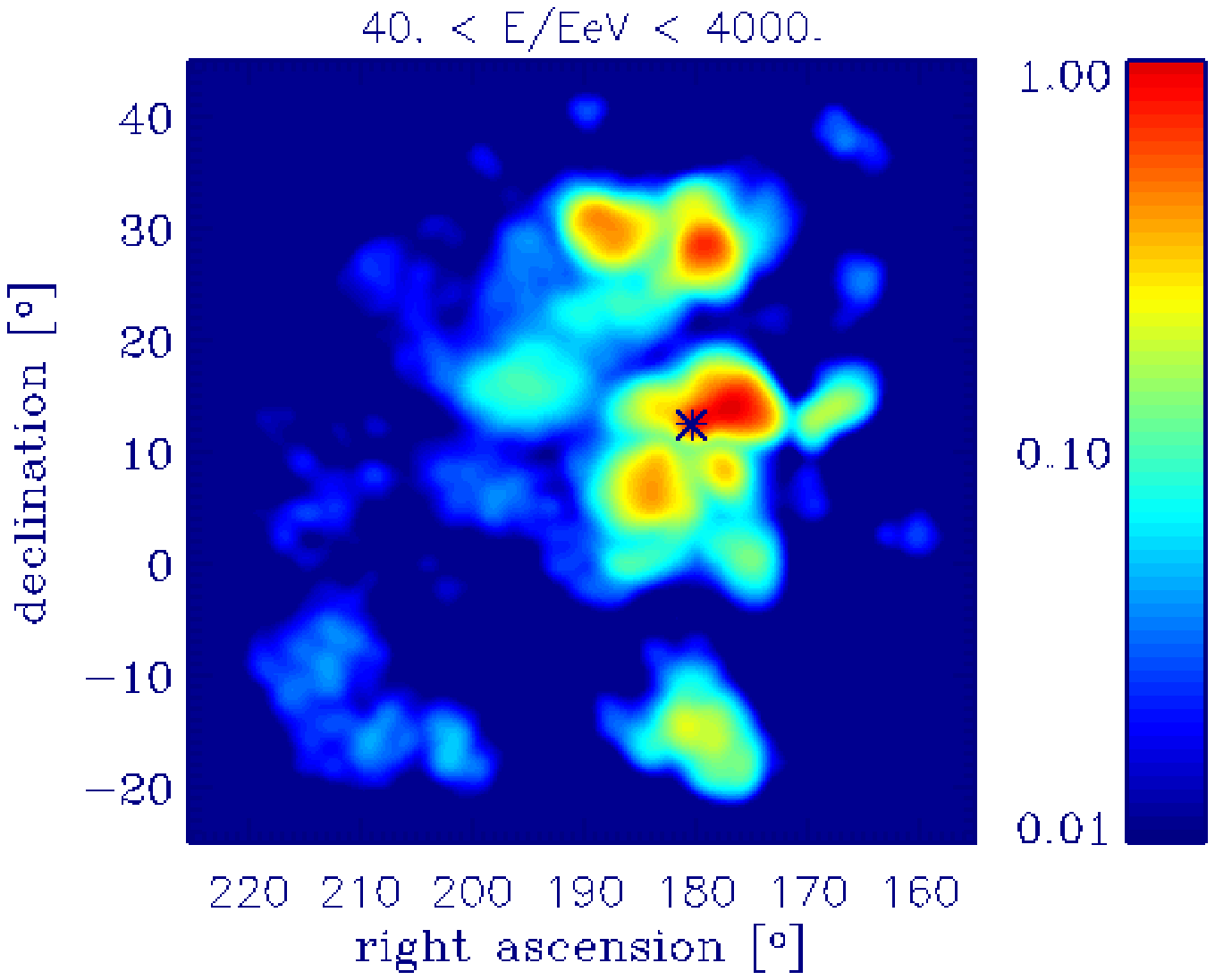}
\includegraphics[width=0.5\textwidth,clip=true]{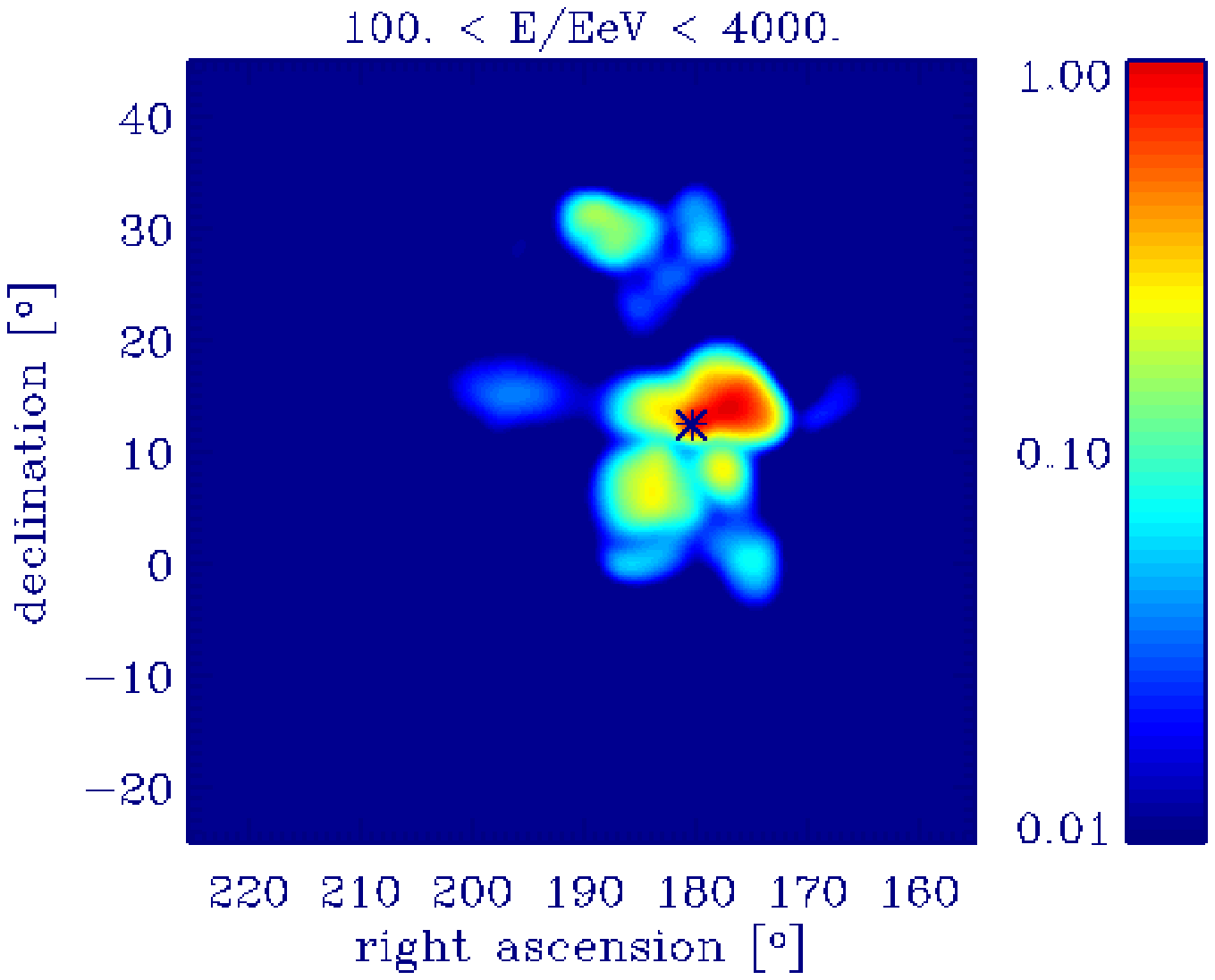}
\caption[...]{Sky distribution of UHECR arrival directions above
$4\times10^{19}\,$eV (left panel) and above $10^{20}\,$eV (right
panel) for proton primaries in scenario ``Virgo''. The source position
is marked by an asterisk. The angular resolution was assumed to
be $1^\circ$.}
\label{fig11}
\end{figure}

Fig.~\ref{fig11} shows an actual sky plot of arrival directions
of events above $4\times10^{19}\,$eV and above $10^{20}\,$eV
for proton primaries in scenario ``Virgo''. Several distinct images
are clearly seen which suggests significant magnetic lensing~\cite{hmrs}
in the EGMF structure from Fig.~\ref{fig1}, right panel.
This is not surprising given
the fact that lensing in general sets in at energies higher
than the transition energy to diffusion which, according to
Eq.~(\ref{Et}), is roughly around $10^{20}\,$eV. The number of
images and the deflection angles decrease with increasing energy,
but are still significant at $10^{20}\,$eV and can be resolved
with sufficient statistics.

\section{Conclusions}
We investigated the impact on cosmic ray observations above $10^{19}\,$eV
of Mpc-scale magnetic fields of $\sim10^{-7}\,$G strength surrounding
ultra-high energy cosmic ray sources for the likely case that magnetic fields
within a few Mpc around Earth are insignificant. We find that
such source fields can strongly modify spectra and composition
at Earth, especially for nearby sources for which the fields can
considerably modify propagation times relative to both energy
loss and photo-disintegration time scales and to the undeflected
propagation time. We found the following generic features:

The spectra are considerably
hardened relative to the injection spectrum at energies below
the usual GZK-like cutoff where energy loss distance and source
distance become comparable. This is caused by an interplay between
diffusion and energy loss: The flux of low energy particles is
suppressed because diffusion spreads them out over a larger volume
due to their much larger energy loss times. This is in contrast
to the case of uniformly distributed magnetic fields which in
general lead to a steepening of the cosmic ray flux below the
GZK cutoff. A hardened sub-GZK spectrum from individual sources
would be consistent with hints of a hard clustered component in
the AGASA data between $10^{19}\,$eV and $10^{20}\,$eV~\cite{teshima1}.

Furthermore, for a nucleus of atomic mass $A$ as injected primary, due
to the kinematics of the photo-disintegration reactions a nucleon
peak appears at energy $\sim E_{\rm max}/A$, where $E_{\rm max}$ is
the maximal nucleus injection energy. This effect is the more prominent
the harder the injection spectrum.
We also found that the details of spectra and composition
depend significantly on the unknown details of the magnetic
fields and the position of the source therein and can thus
not be predicted.

Further cutoffs towards low energies can be induced if
the source is active only since a time smaller than the
typical delay time at this energy. Obviously, the characteristics of
such features also depend on unknown details of the source.

Our simulations finally show that even for iron primaries,
extra-galactic magnetic fields from large scale structure
simulations are not strong and extended enough to explain the
observed large scale isotropy of ultra high energy cosmic ray
arrival directions in terms of a single nearby source. This
would require more homogeneous fields such as in Ref.~\cite{agrs}.

Next generation experiments such as the Pierre Auger
Observatories~\cite{auger} and the EUSO project~\cite{euso}
will accumulate sufficient statistics to establish
spectra and distributions of composition and arrival directions
from individual sources. A potentially strong influence of magnetic
fields surrounding individual sources should thus be kept in mind when
interpreting data from these experiments. This is true even
if ultra-high energy cosmic rays arrive within a few degrees
from the source position.

\ack
This work partly builds on earlier collaborations with
Gianfranco Bertone, Torsten En\ss lin, Claudia Isola,
Martin Lemoine, and Francesco Miniati. I am grateful to
Francesco Miniati for comments on the manuscript.
Luis Anchordoqui and Haim Goldberg
are acknowledged for discussions on propagation of nuclei in
magnetic fields. I thank the Max-Planck Institut f\"ur Astrophysik
where part of this work was done for hospitality.


\end{document}